\documentclass
{mn2e}
\usepackage{epsfig}
\usepackage{amsmath, amssymb,bm}
\usepackage{psfrag}

\newcommand{\msun}{\rm M_{\odot}}

\newcommand{\me}{\dot M_{\rm Edd}}
\newcommand{\md}{\dot M_{\rm dyn}}
\newcommand{\mo}{\dot M_{\rm out}}
\newcommand{\mw}{\dot M_{\rm w}}
\newcommand{\led}{L_{\rm Edd}}
\newcommand{\rph}{R_{\rm ph}}
\def \msun{\rm M_{\odot}}
\def \me{{\dot M_{\rm Edd}}}
\def \macc{{\dot m_{\rm acc}}}
\def \mout{{\dot m_{\rm w}}}
\def \moutt{{\dot m_{\rm w_{10}}}}
 
\begin{document}
%%%%%%%%%%%%%%%%%%%%%%%%%%%%%%%%%%%%%%%%%%%%%%%%%%%%%%%%%%%%%%%%%%%%%%%%%%%%%%
%% Title Details and Page Header                                            %%
%%%%%%%%%%%%%%%%%%%%%%%%%%%%%%%%%%%%%%%%%%%%%%%%%%%%%%%%%%%%%%%%%%%%%%%%%%%%%%
\title[Black Hole Winds II]{Black Hole Winds II: Hyper--Eddington Winds and Feedback}

\author[Andrew King \& Stuart I. Muldrew] 
{
\parbox{5in}{Andrew King$^{1, 2}$ \& Stuart I. Muldrew$^{1}$ }
\vspace{0.1in} 
\\ $^1$ Department of Physics \& Astronomy, University
of Leicester, Leicester LE1 7RH UK\\
$^2$ Astronomical Institute Anton Pannekoek, University of Amsterdam,
Science Park 904, 1098 XH Amsterdam, Netherlands}
\date{Accepted 2015 October 6.  Received 2015 September 30; in original form 2015 August 20}

\maketitle

%%%%%%%%%%%%%%%%%%%%%%%%%%%%%%%%%%%%%%%%%%%%%%%%%%%%%%%%%%%%%%%%%%%%%%%%%%%%%%
%% Abstract, Keywords and contact details                                   %%
%%%%%%%%%%%%%%%%%%%%%%%%%%%%%%%%%%%%%%%%%%%%%%%%%%%%%%%%%%%%%%%%%%%%%%%%%%%%%%
\begin{abstract}
We show that black holes supplied with mass at hyper--Eddington rates drive outflows with mildly sub--relativistic velocities. These are $\sim 0.1 - 0.2c$ for Eddington accretion factors $\macc \sim 10 - 100$, and $\sim 1500\,{\rm km\, s^{-1}}$ for $\macc \sim 10^4$. Winds like this are seen in the X--ray spectra of ultraluminous sources (ULXs), strongly supporting the view that ULXs are stellar--mass compact binaries in hyper--Eddington accretion states. SS433 appears to be an extreme ULX system ($\macc \sim 10^4$) viewed from outside the main X--ray emission cone. For less extreme Eddington factors $\macc \sim 10 - 100$ the photospheric temperatures of the winds are $\sim 100$\, eV, consistent with the picture that the ultraluminous supersoft sources (ULSs) are ULXs seen outside the medium--energy X--ray beam, unifying the ULX/ULS populations and SS433 (actually a ULS but with photospheric emission too soft to detect). For supermassive black holes (SMBHs), feedback from hyper--Eddington accretion is significantly more powerful than the usual near--Eddington (`UFO') case, and if realised in nature would imply $M - \sigma$ masses noticeably smaller than observed. We suggest that the likely warping of the accretion disc in such cases may lead to much of the disc mass being expelled, severely reducing the incidence of such strong feedback. We show that hyper--Eddington feedback from bright ULXs can have major effects on their host galaxies. This is likely to have important consequences for the formation and survival of small galaxies. 
\end{abstract} 


\footnotetext[1]{E-mail: ark@astro.le.ac.uk}

%%%%%%%%%%%%%%%%%%%%%%%%%%%%%%%%%%%%%%%%%%%%%%%%%%%%%%%%%%%%%%%%%%%%%%%%%%%%%%
%% Introduction                                                             %%
%%%%%%%%%%%%%%%%%%%%%%%%%%%%%%%%%%%%%%%%%%%%%%%%%%%%%%%%%%%%%%%%%%%%%%%%%%%%%%
\section{Introduction}
\label{intro}
It is now widely accepted that winds driven by accretion on to black holes have major effects on their surroundings. Wide--angle winds from supermassive holes (SMBH) in galaxy centres are probably the origin of the SMBH -- galaxy scaling relations (for reviews of observations and theory see Kormendy \& Ho, 2013, and King \& Pounds 2015). King \& Pounds (2003) gave a simple picture of black hole winds for the case when accretion on to the black hole is close to the Eddington value. This paper extends this to cover cases where the hole's mass supply rate is much larger. 

Shakura \& Sunyaev (1973) set out a general theoretical picture of disc winds. They considered accretion discs where the mass supply at the outer edge would be super--Eddington near the black hole. At large distances from the hole, gas spirals inwards through an accretion disc. Radiation pressure becomes significant in the disc at the spherization radius $R_{\rm sph}$,  where the local energy release $GM\dot M/R_{\rm sph}$ is of order the Eddington luminosity
\begin{equation}
\led = \frac{4\pi GMc}{\kappa}
\label{ledd}
\end {equation}
with $M$ the black hole mass, $\dot M$ the instantaneous accretion rate, and $\kappa$ the electron scattering opacity. Some of the accreting gas is driven out as a wind by radiation pressure, and Shakura \& Sunyaev assumed that this occurs at all disc radii $R < R_{\rm sph}$ in such a way that $GM\dot M(R)/R \sim \led$ at all $R$. This means that the accretion rate through the disc decreases as $\dot M(R) \propto R$, reaching the Eddington rate $\me = \led/\eta c^2$  (with $\eta \sim 0.1$  the accretion efficiency) at the last stable circular orbit, and so growing the black hole mass at the rate $\me$. The total radiation output is
\begin{equation}
L \simeq l\led,
\label{ledd2}
\end{equation}
with
\begin{equation}
l \simeq [1 + \ln\macc]
\label{ledd3}
\end{equation}
and $\macc = \dot M/\me$ is the Eddington accretion ratio between mass supply and $\me$ (the logarithmic factor is sometimes taken as $\ln(1+\macc)$).  Numerical simulations (Ohsuga, 2009; Sadowski et al., 2014;  Jiang et al., 2014) qualitatively confirm much of this simple picture, but generally find that the central accretion rate can be somewhat higher than $\me$.

The resulting configuration is a quasispherical wind removing the excess mass from the disc, with a near--vacuum along the rotation axis because of the combined effect of centrifugal repulsion and radiation pressure (see Fig. 1). Much of the radiation escapes preferentially through these evacuated funnels, but a component of luminosity $\sim \led$ is radiated almost isotropically from the near--spherical photosphere of the wind.

Wind feedback of this type is inherently more effective than that from jets (see Fabian, 2012, for a review), which require some means of spreading their effects more widely (e.g. changes of direction) to influence galaxies as strongly.

The SMBH winds thought to power feedback in galaxies are now known as UFOs (`ultrafast outflows'; cf Tombesi et al. 2010a,b: see King \& Pounds 2015 for an overview). They are characterised by terminal velocities $v \sim 0.1c$, and mass rates $\mw$ implying  a thrust scalar close to the Eddington value, i.e.
\begin{equation}
\dot P_w =\dot M_{\rm w}v \simeq {\led\over c}.
\label{ufo}
\end{equation}
King \& Pounds (2003) pointed out that
for a black hole fed at a rate with Eddington accretion ratio $\dot m \sim 1$, the low optical depth $\tau \simeq 1$ of the radiation--pressure--driven wind would allow most of the photons produced by accretion to escape the wind after about one scattering. The front--back symmetry of electron scattering then implies that on average each photon gives up all its momentum to the wind. This gives the relation (\ref{ufo}), which characterises this single--scattering limit as producing a wind whose momentum is very close to that of the driving radiation field.

The basic relation
(\ref{ufo}) implies 
\begin{equation}
v = {\eta c\over \mout} \simeq 0.1c,
\label{vufo}
\end{equation}
as observed. It follows that the mechanical luminosity $L_w$ of the black hole wind in the single--scattering limit is 
\begin{equation}
L_w = {1\over 2}\mw v^2 = \frac{\eta^2 c^2\me}{ 2\mout} \simeq {\eta\over 2}\led \simeq 0.05\led
\label{lufo},
\end{equation}
highlighting the relatively weak coupling of photons and matter when $\macc\sim \mw \sim 1$, and so $\tau \sim 1$. Numerical simulations of mildly super--Eddington winds (e.g. Ohsuga \& Mineshige, 2011 -- their Models A and B) agree very well with these predictions.

This simple picture is remarkably successful in describing SMBH feedback. Equations ({\ref{ufo}}, {\ref{vufo}}) are in explicit agreement with observations of UFOs, while galaxy--scale molecular ouflows driven by the central AGN have mechanical luminosities close to (\ref{lufo}) (Cicone et al., 2014). There is even the direct example of a UFO observed in the centre of an AGN--driven molecular outflow with precisely the expected relations ({\ref{ufo}}, {\ref{vufo}}, \ref{lufo}) (Tombesi et al., 2015). Theoretically, this UFO picture is a reasonable description of most sustained SMBH accretion in galaxies because the long--term accretion rates are rarely super--Eddington. Even the most extreme estimate of gas accretion in a galaxy bulge with velocity dispersion $\sigma$ and gas fraction $f_g$, i.e. the dynamical rate $\md = f_g\sigma^3/G$, is not very super--Eddington ($\macc \la 60$) for an SMBH mass close to the $M - \sigma$ relation (King, 2007; King \& Pounds 2015). 

But at other mass scales there are many systems where the mass supply must be strongly super--Eddington, and the single--scattering UFO limit cannot hold. For example, a predominantly radiative star filling its Roche lobe in a binary system transfers much of its mass ($M_2$) on its thermal timescale to the companion (mass $M_1$) if $M_2 \ga M_1$. Phases like this occur often in binary evolution. They are unavoidable for example once the donor star in a high--mass X--ray binary fills its Roche lobe, and can give Eddington ratios as high as $\macc \ga 10^4$. This probably holds in the precessing--jet binary system SS433 (King, Taam \& Begelman, 2000).
It appears very likely that hyper--Eddington mass rates are the essential property characterising ultraluminous X--ray sources (ULXs)\footnote{These rates would also be super--Eddington for neutron stars or white dwarfs, so the recent discovery of regular X--ray pulsing from a  ULX (Bachetti et al., 2014), indicating a neutron--star accretor, is in line with theoretical expectations (Fabbiano et al., 2003; King, 2009).
For brevity, we nevertheless refer to the accretor as a `black hole' throughout.}. Evidently we cannot expect the UFO approximation $\tau \sim 1$ to hold for such high--density accretion flows, so this paper investigates the case $\tau \gg 1$.

\section{Hyper--Eddington Black Hole Winds}
\label{hyperedd}
For black hole accretion in the case $\macc \gg 1$ we abandon the single--scattering relation (\ref{ufo}) and instead
assume (and justify below) that the resulting wind is so optically thick that the photon field couples very strongly with it, and gives the wind a mechanical luminosity 
$L'_w$ comparable with that of the radiation field, i.e. $L'_w \simeq l\led$. For hyper--Eddington accretion this relation replaces the equation $\dot P_w \simeq \led/c$ defining the single--scattering case. Numerical simulations by Hashizume et al. (2015) produce just this kind of result:
matter is blown away in almost all directions outside the equatorial plane, with mechanical luminosity comparable to the photon luminosity.

We will see below that the wind is launched at disc radii several times the gravitational radius  $R_g = GM/c^2$. At these larger radii the accretion luminosity available to drive the wind is $\sim l'\led$ 
with $l' \la l$, somewhat less than the full luminosity $l\led$ (cf equation (\ref{ledd3})),  as some of the photons from the inner regions $R \sim R_g$ find the open funnels around the rotation axis and escape that way, rather than contributing to wind driving. Accordingly we take the defining relation for hyper--Eddington winds as
\begin{equation}
L'_w = \frac{1}{2}\mw v'^2 \simeq l'\led = \frac{2l'}{\eta}L_w,
\label{hyper}
\end{equation}
rather than (\ref{ufo}), which defines the single--scattering case. Then hyper--Eddington winds have the velocity
\begin{equation}
v'\simeq \left(\frac{2l'\eta}{\mout}\right)^{1/2}c.
\label{vhyper}
\end{equation}
Fig. 2 shows $v'/c$ as a function of $\mout$.

\begin{figure}
  \includegraphics[width=86mm]{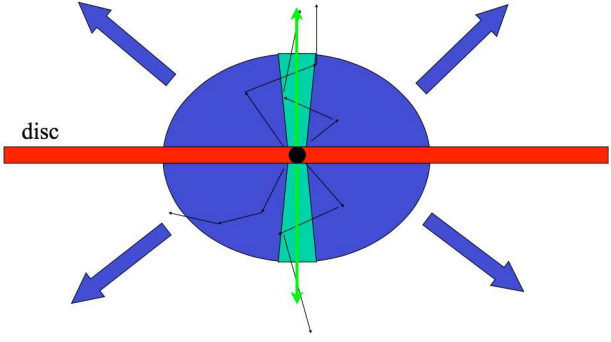}
  \caption{Schematic picture of the central region of a super--Eddington source. The plane of the central part of the accretion disc is assumed to have aligned in the spin plane of the black hole via the Lense--Thirring effect (cf King et al., 2005). The excess accretion is expelled in a quasispherical wind, with evacuated emission cones around the rotational axis. Much of the accretion luminosity $l\led$ escapes through these cones (photon tracks shown in black), while a significant component $l' \led \ga \led$ is radiated from the wind photosphere. The green arrows indicate the directions of jets, if present. An observer viewing the source from within one of the emission cones would assign it an apparent luminosity $\gg l\led$ if it is assumed to radiate isotropically, and so identify it as an ultraluminous X--ray source (ULX). Strongly blueshifted X--ray absorption lines are also visible in these directions. Observers viewing from outside these cones would identify the wind photospheric emission as an ultraluminous supersoft source (ULS). SS433 is viewed in this direction, but its extreme Eddington factor $\macc \sim 10^4$ makes its photospheric emission too soft to be detected.}
\label{fig1}
\end{figure}

We note that for Eddington factors $10 < \mout < 100$
we get $v\sim 0.1- 0.2c$, similar to UFOs, despite the very different physical regime. Observations of ULX absorption lines reveal winds with velocities of order $0.1c - 0.2c$ (Middleton et al., 2014) in agreement with equation (\ref{vhyper})  for $\mout \sim 10 - 100$. In SS433 the much larger Eddington factor $\mout \sim 10^4$ implies a quasispherical wind speed close to the observed velocity width $\sim 1500\,{\rm km\, s^{-1}}$  of the so--called `stationary' H--alpha line. Recent numerical simulations (Jiao et al., 2015) also find winds with velocities $\sim 0.1 - 0.2c$
except very close to the rotational axis, where the velocity is higher, as expected.

To check self--consistency we estimate the optical depth of a wind with velocity (\ref{vhyper}). A quasispherical wind with mass rate $\mw$  and velocity $v'$ has density
\begin{equation}
\rho(R) = \frac{\mw}{4\pi R^2 v'}
\label{density}
\end{equation}
and so optical depth 
\begin{equation}
\tau(R_0) = \int_{R_0}^{\infty}\frac{\mw\kappa}{4\pi R^2 v'}{\rm d}R 
= \frac{\mw\kappa}{4\pi v' R_0}
\label{tauR}
\end{equation}
at any given radius $R_0$. The wind must be launched from radii where the local escape velocity is of order $v'$, i.e. from $R_{\rm wind} \sim 2GM/v'^2$, so the optical depth through it is
\begin{equation}
\tau_{\rm wind} \sim \frac{\mw\kappa v'}{8\pi GM} = \frac{\mw\kappa}{8\pi GM}\left(\frac{2l'\eta }{\mout}\right)^{1/2}c,
\end{equation}
where we have used (\ref{vhyper}) at the last step. Now using (\ref{ledd}, \ref{ledd2}, \ref{vhyper}) we get finally
\begin{equation}
\tau_{\rm wind} \sim \left(\frac{l'\mout }{2\eta}\right)^{1/2} = \frac{l'c}{v'}
\label{taul}
\end{equation}
which gives $\tau_{\rm wind} \sim10$ for  $\mout \sim 10 - 100$, and $\tau \gg 1$ for larger $\dot m$. This justifies the assumption that matter and radiation are well--coupled if $\macc \sim \mout\gg 1$, expressed as equation (\ref{hyper}). 

\begin{figure}
  \psfrag{x}[][][1][0]{${\rm log}_{10}\mout$}
  \psfrag{y}[][][1][0]{$v'/c$}
  \psfrag{a}[l][][1][0]{$\eta=0.1$}
  \psfrag{b}[l][][1][0]{$\eta=0.4$}
  \includegraphics[width=86mm]{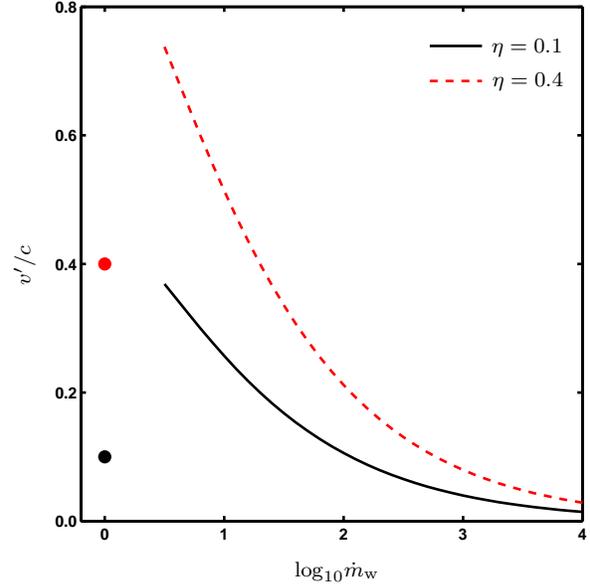}
  \caption{The expected speed $v'$ of a hyper--Eddington black hole wind as a function of Eddington factor $\mout$ for the cases $\eta = 0.1$ (modest black hole spin) and $\eta = 0.4$ (high black hole spin). We take $l' \simeq l$ (cf eqn \ref{ledd3}). The expected values of $v$
for an Eddington wind (UFO) with $\mout$ are shown (heavy dots) in these two cases for comparison.}
\label{fig2}
\end{figure}

An important feature of these hyper--Eddington winds is their much greater momentum rate $\dot P'_w$ compared with UFOs:
\begin{equation}
\dot P'_w = \mw v' = \frac{2L'_w}{v'} = \frac{2l'c}{v'}\frac{\led}{c} =\frac{2l'c}{v'}\dot P_w
\label{hypermom}
\end{equation}
(cf equation (\ref{ufo})). 

\section{Observational Appearance of Hyper--Eddington Sources}
We expect a hyper--Eddington system to radiate most of its accretion luminosity $l\led$ from the evacuated funnels which form a narrow double cone around the rotational axis (see Fig. 1). 
The resulting collimation (cf King, 2009) makes the source appear as a ULX, with apparent luminosity $\gg l\led$ for an observer viewing the system in this direction but assuming that its flux is isotropic. 

The source must also radiate a luminosity $\sim \led$ from the quasispherical wind photosphere.
The optical depth defining this photosphere is determined by the {\it effective} absorption
coefficient $\mu_{\rm eff} = (\mu_{\rm es}\mu_{\rm ff})^{1/2}$, where 
\begin{equation}
\mu_{\rm es} = 6.7\times 10^{-25}N\,{\rm cm}^{-1}
\label{es}
\end{equation}
and
\begin{equation}
\mu_{\rm ff} \simeq 5\times 10^8\nu^{-3}N^2T^{-1/2}\,{\rm cm}^{-1}
\label{ff}
\end{equation}
are the opacity coefficients for electron scattering and free--free absorption respectively (e.g. Shakura \& Sunyaev 1973), where $N\simeq \rho/m_p$ is the number density of the wind, with $m_p$ the proton mass, $\nu$ is the radiation frequency and $T$ the wind temperature.  

In practice the steep dependences $R _0\propto T_0^{-2}, N_0 \propto R_0^{-2} \propto T_0^{-4}$ of the photospheric radius $R_0$, number density $N_0$ and temperature $T_0$ (cf eqn \ref{density}) mean that it is always close to the pseudophotospheric radius $R_{\rm ph}$ for electron scattering, defined simply by setting $\tau(R_{\rm ph}) = 1$ in (\ref{tauR}), since $\mu_{\rm eff} \simeq \mu_{\rm es}$ there, as we demonstrate here. 

Defining $\rph$  as above, and using (\ref{hyper}), i.e
\begin{equation}
\rph = \frac{\mo\kappa}{4\pi v'} = \frac{\kappa l' \led}{2\pi v'^3} = \frac{2l'GMc}{v'^3} = 2l'\left(\frac{c}{v'}\right)^3R_g  
\label{rph}
\end{equation}
we find
\begin{equation}
\rph \simeq 1\times 10^{4}l'^{-1/2}\moutt^{3/2}\eta_{10}^{-3/2}M_{10}\, {\rm km}  
\label{rph2}
\end{equation}
where $\moutt = \mout/10, \eta_{0.1} = \eta/0.1$ and $M_{10} = M/10\msun$.
The effective temperature of this pseudophotosphere is given by
\begin{equation}
T_{\rm eff}^4 \simeq \frac{\led}{4\pi\sigma \rph^2},
\end{equation}
with $\sigma$ the Stefan--Boltzmann constant, which gives
\begin{equation}
T_{\rm eff} = 1\times 10^6l'^{1/4}\moutt^{-3/4}\eta_{0.1}^{3/4}M_{10}^{-1/4}\, {\rm K}
\label{teff}
\end{equation}
or 
\begin{equation}
kT_{\rm eff} = 100l'^{1/4}\moutt^{-3/4}\eta_{0.1}^{3/4}M_{10}^{-1/4}\, {\rm eV}.
\label{teff2}
\end{equation}

Now we can evaluate $\mu_{\rm es}, \mu_{\rm ff}$ at this point, substituting $T = T_{\rm eff} \simeq 10^6\,{\rm K}, N = N_0 \simeq 10^{24}\rph \simeq 10^{18}\,{\rm cm}^{-3}$ and $\nu \simeq kT_{\rm eff}/h \simeq 10^{16}\,{\rm Hz}$ to find
\begin{equation}
\mu_{\rm eff} \simeq 0.84\mu_{\rm es}
\end{equation}
This shows that in practice the wind photosphere is alway close to the scattering pseudophotosphere, so to a good approximation
we can take eqns (\ref{rph2}, \ref{teff2}) as defining the effective photosphere of the wind. This
immediately tells us that stellar--mass hyper--Eddington sources have precisely the defining properties of the ultraluminous supersoft sources (ULSs): soft thermal spectra with photospheric radii $\sim 10^4$\, km and effective temperatures $\sim 100$\, eV (Kong \& di Stefano, 2003; see Urquhart \& Soria, 2015 for a recent summary). Evidently ULSs are hyper--accreting stellar--mass black holes or neutron stars, just as ULXs appear to be. The only distinction between ULXs and ULSs is the viewing angle of the observer. In ULXs we see the system from inside the  relatively narrow emission cone of medium--energy X--rays along the accretion disc's rotational axis (see Fig. 1), while in ULSs we are outside this cone and see only the supersoft emission specified by (\ref{teff}, \ref{teff2}). The lack of medium--energy X--ray emission from ULSs makes it hard to detect their black hole winds, as these cannot produce the blueshifted X--ray absorptions of hydrogen-- and helium--like iron which characterize ULX winds (Middleton et al., 2014). If the wind had sufficiently low ionization it might be possible to detect blueshifted lines at lower excitation, but this is unclear.

We note that in SS433 ($\mout \simeq 10^4$) we have $kT_{\rm eff} \sim 10$\, eV, which is too soft to be detectable given this system's heavy reddening. So this extreme system is observationally neither a ULX nor a ULS, although in terms of viewing angle it is in the latter group. It is picked out only because of the very strong periodic red-- and blue--shifts of the H--alpha lines emitted by its precessing jets, which are evidently related to its extreme Eddington factor.

\section{Feedback}

As we remarked at the start of this paper, an important property of black hole winds is their ability to affect their surroundings by injecting energy and momentum into them. The nature of this interaction is fixed by the physics of the shocks where the winds impact the surrounding gas. If these shock are efficiently cooled we have momentum--driven flow, and only ram pressure is communicated to the ambient medium. If instead the shocks do not cool we have energy--driven flow, and all of the wind energy is used to drive an adiabatically expanding bubble into the surroundings. 

This duality is the basic reason that SMBH in the centres of galaxies first grow to the $M - \sigma$ mass, and then expel most of the interstellar gas from the central bulge, limiting their further growth (see King \& Pounds 2015 for a review).  When $M$ is below the $M - \sigma$ value the feedback shocks are cooled by the inverse Compton effect, remaining in the momentum--driven regime and affecting the host ISM only gently. But once this mass is reached, Compton cooling becomes inefficient. Feedback is then in the energy--driven regime, expelling the bulge gas on a dynamical timescale and preventing SMBH mass growth beyond $M -\sigma$.

In the single--scattering UFO limit appropriate to SMBH accretion there is a fairly coherent picture of how feedback works. We briefly summarize this here, before going on to show how things change for hyper--Eddington winds.

\subsection{Feedback from Eddington (UFO) Winds}

King \& Pounds (2015) give a recent summary of this case, which describes SMBH interactions with the host galaxy spheroid (characterised by velocity dispersion $\sigma$ and gas fraction $f_g \sim 0.1$). The radiation field of the accreting SMBH is far cooler than the UFO wind shocks against the bulge gas. If these are within the cooling radius
\begin{equation}
R_C \simeq 500M_8^{1/2}\sigma_{200}\,{\rm pc},
\label{ufocool}
\end{equation}
where $M_8 = M/10^8\msun$ and $\sigma_{200} = \sigma/200\,{\rm km\, s^{-1}}$, the radiation field
is dense enough that the inverse Compton effect can efficiently cool the shocks and enforce momentum--driven flow.  Then feedback is a contest between 
the momentum flux $\dot P_w = \mw v \simeq \led/c$ carried by the UFO wind, and the weight $4f_g\sigma^4/G$ of the bulge gas. The two balance when the SMBH mass takes the value
\begin{equation}
M = M_{\sigma} = {f_g\kappa\over \pi G^2}\sigma^4 \simeq 3\times 10^8\msun\sigma_{200}^4.
\label{msig}
\end{equation}
For $M < M_{\sigma}$, UFO feedback tries to sweep up the bulge gas into a shell, but this becomes too heavy and falls back. So for such black hole masses, the SMBH has no effect on the gas potentially feeding its growth. 

But once $M > M_{\sigma}$, the swept--up gas is pushed away from the SMBH to radii $\gg R_C$ where the radiation field is too dilute to cool the shocks efficiently. Instead the shocked wind now forms an adiabatically expanding hot bubble, injecting all the wind luminosity $L_w =(\eta/2)\led$ (from eqn \ref{lufo}) into expansion.  This expels all the gas from the galaxy bulge, and so halts SMBH mass growth with only a very slight increase $\Delta M$ of SMBH mass beyond $M_{\sigma}$: the 
gas binding energy is of order
\begin{equation}
 E_g \sim f_gM_b\sigma^2 \sim 8\times 10^{57}M_{11}\sigma_{200}^2\,{\rm erg},
\label{gasbind}
\end{equation}
where $M_b = 10^{11}\msun M_{11}$ is the bulge stellar mass, while the wind energy produced by accreting a mass $\Delta M$ to the SMBH is of order
\begin{equation}
E_{\rm acc} \sim \frac{\eta}{2}\eta c^2 \Delta M \sim10^{60}\frac{\Delta M}{ M}M_8\,{\rm erg},
\label{bhbind}
\end{equation}
with $M = 10^8\msun M_8$. So only a very small fractional increase
\begin{equation}
\frac{\Delta M}{M} \sim 8\times 10^{-3}\frac{M_{11}}{M_8}\sigma_{200}^2
\label{unbind}
\end{equation}
of the SMBH mass is needed to remove all the gas.

\begin{figure}
  \psfrag{x}[][][1][0]{${\rm log}_{10}\mout$}
  \psfrag{y}[][][1][0]{$M_{\sigma}'/M_{\sigma}$}
  \psfrag{a}[l][][1][0]{$\eta=0.1$}
  \psfrag{b}[l][][1][0]{$\eta=0.4$}
  \includegraphics[width=86mm]{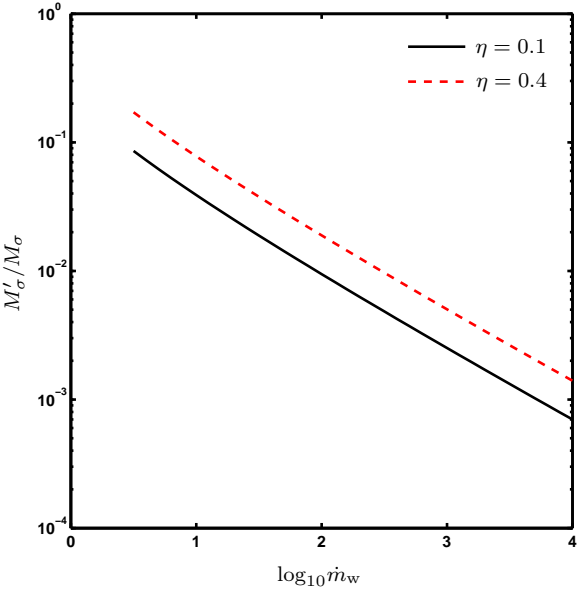}
    \caption{Predicted $M - \sigma$ mass $M_{\sigma}'$
     as function of Eddington ratio $\mout$, 
    relative to the value $M_{\sigma}$ [eq \ref{msig}] given by a UFO wind (with $\mout\simeq 1$).}
\label{fig3}
\end{figure}

\begin{figure}
  \psfrag{x}[][][1][0]{${\rm log}_{10}\mout$}
  \psfrag{y}[][][1][0]{$v_{\rm out}'/v_{\rm out}~,~\dot M_{\rm out}'/\dot M_{\rm out}$}
  \psfrag{a}[l][][1][0]{$\eta=0.1$}
  \psfrag{b}[l][][1][0]{$\eta=0.4$}
  \includegraphics[width=86mm]{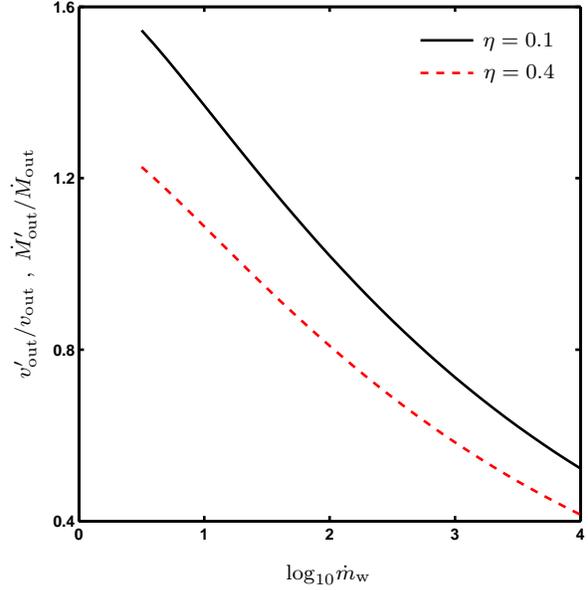}
  \label{fig4}
\caption{Variation of the large--scale (typically molecular) outflow velocity [eq.\ref{vprimeout}] and mass rate [eq \ref{mdotoutprime}] driven by black--hole winds with Eddington factor 
$\mout$, relative to the values [eq \ref{vout}], [eq \ref{mdotout}] driven by a UFO wind (with 
$\mout \simeq 1$). }
\end{figure}

The dynamics of the expansion (Zubovas \& King, 2012) show that the gas is expelled with velocity 
\begin{equation}
v_{\rm out} = \frac{4}{3}\left[\frac{L_w(M_\sigma) l G}{3f_g\sigma^2}\right]^{1/3} = 1230\sigma_{200}^{2/3}l^{1/3}\,{\rm km\,s^{-1}},
\label{vout}
\end{equation}
(where $L_w(M_\sigma)$ is $L_w$ evaluated at BH mass $M = M_{\sigma}$) and mass rate
\begin{equation}
\dot M_{\rm out} = \frac{2f_g\sigma^2}{G}v_{\rm out} = 4060\sigma_{200}^{8/3}\msun\,{\rm yr}^{-1}
\label{mdotout}
\end{equation}
Observations of AGN--driven molecular outflows (e.g. Cicone et al., 2014 and references therein, and Tombesi et al., 2015) agree very well with the predictions (\ref{vout}, \ref{mdotout}). Note that the molecular outflow quantities $v_{\rm out}, \dot M_{\rm out}$ are quite distinct from the {\it black hole wind} velocity $v$ and outflow rate $\mw$, with 
$v \sim 0.1c \gg v_{\rm out}$ and $\mw \ll \dot M_{\rm out}$. The observations of Tombesi et al. (2015) report both of these distinct velocities and mass rates.

\subsection{Feedback from hyper--Eddington winds}

For the hyper--Eddington case we can repeat the steps of the previous sub-section, substituting $\dot P'_w, L'_w$ for $\dot P_w, L_w$. We retain the same parametrizations ($\sigma_{200}$ etc) to make comparison straightforward, but discuss later the consequences for stellar--mass black hole systems in dwarf galaxies.
The cooling radius $R_C$ remains unchanged from the Eddington case, but the greater thrust of a hyper--Eddington wind gives a modified  $M - \sigma$ mass
\begin{equation}
M'_{\sigma} = \frac{v'}{2l'c}M_{\sigma}
\simeq 3\times 10^8\msun\sigma_{200}^4\left(\frac{\eta}{2{\it l'\mout}}\right)^{1/2}
\label{msigprime}
\end{equation}
Figure 3 shows how $M'_{\sigma}$ varies with $\mout$. Here the Eddington factor potentially has a significant effect, which we consider in the next Section.

The bulge gas binding energy $E_g$ is of course the same as in the UFO case, but the wind energy gain from a SMBH mass increase $\Delta M'$ is now
\begin{equation}
E_{\rm acc} \sim \eta l'c^2 \Delta M' \sim 2\times10^{61}\frac{l'\Delta M'}{ M}M_8\,{\rm erg},
\label{bhbindprime}
\end{equation}
so the fractional SMBH mass increase needed to remove all the gas is
\begin{equation}
\frac{\Delta M'}{M} \sim 4\times 10^{-4}\frac{M_{11}}{l'M_8}
\sigma_{200}^2
\label{unbindprime}
\end{equation}
For BH masses $> M'_{\sigma}$, gas is expelled in an energy--driven flow
with the modified velocity
\begin{equation}
 v'_{\rm out} = \frac{4}{3}\left[\frac{L'_w(M'_\sigma) l' G}{3f_g\sigma^2}\right]^{1/3}
= \left(\frac{v'}{\eta c}\right)^{1/3}v_{\rm out}
\end{equation}
so that
\begin{equation}
v'_{\rm out}=v_{\rm out}
\left(\frac{2l'}{\eta\mout}\right)^{1/6},
\label{vprimeout}
\end{equation}
and the mass rate
\begin{equation}
\dot M'_{\rm out} = \frac{2f_g\sigma^2}{G}v'_{\rm out} = \dot M_{\rm out}
\left(\frac{2{\it l'}}{\eta\mout}\right)^{1/6}.
\label{mdotoutprime}
\end{equation}
Thus even for extreme values of $\mout$, the large--scale outflow velocity $v'_{\rm out}$ and mass rate $\dot M'_{\rm out}$ hardly differ from the values for $\mout \simeq 1$ (see Figure 4).

\section{Discussion}

For supermassive black hole accretion, we see from Figs 2, 3 and 4 that the differences 
between the UFO and hyper--Eddington cases are quite small, with the important exception of the $M - \sigma$ mass. Strongly super--Eddington accretion would produce a much lower final SMBH mass than observed. The SMBH mass is hard to determine in most galaxies, and harder still if it is small, so it is quite possible that a second $M - \sigma$ relation like this might exist, parallel to but below the current one (and presumably with large scatter unless something makes all the Eddington factor very similar). However since a low SMBH mass makes $\dot m$ larger for a given mass supply rate, we have to explain why the currently--observed $M - \sigma$ relation lies where it does, i.e. corresponding to $\mout \simeq 1$ -- an initially strongly undermassive SMBH might self--consistently keep $\mout\gg 1$ and so lead to small final masses $M_{\sigma}'$. This must mean that in many galaxies strongly super--Eddington accretion on to the central SMBH occurs only rarely, or not at all.

A possible answer which we shall explore in a separate paper uses the idea the accretion disc in SMBH accretion is very likely to be warped, since accretion events do not appear to correlate with the SMBH spin direction, as revealed by AGN jet directions. Accretion may well be via a series of randomly--oriented discs of given mass, cf the papers by Sanders (1981) and the `chaotic accretion' picture of King \& Pringle (2006, 2007; see also King \& Nixon, 2015). In this case, it seems possible that if central accretion reached super--Eddington rates, the powerful black hole wind could sweep away the warped outer regions of the disc, which contain most of the mass of any event. This would then tend to suppress hyper--Eddington accretion, and so the tendency towards low--mass $M -\sigma$ masses.

In stellar--mass cases we know from observation that sustained hyper--Eddington accretion can occur, as in SS433, consistent with the fact that the binary companion continuously refills the disc. The interesting question here is under what conditions hyperaccretion makes feedback from stellar--mass binaries significant in a dwarf galaxy, since ULXs in such galaxies are sometimes observed have very strong effects on their hosts (e.g. Pakull \& Mirioni, 2003), producing nebulae which occupy a significant fraction of the galaxy. A simple test of this idea is to ask what the value of $M'_{\sigma}$ is for a dwarf galaxy (i.e. one with $\sigma \simeq 10\,{\rm km\,s^{-1}}$), as a function of Eddington parameter $\mout$. If $M'_{\sigma}$  can take stellar values for reasonable $\mout$ we would expect a ULX with the right mass and Eddington ratio to create a large nebula.
From (\ref{msigprime}) we find
\begin{equation}
M'_{\sigma} 
\simeq 1.9\times 10^3\msun\sigma_{10}^4\left(\frac{\eta}{2{\it l'\mout}}\right)^{1/2}
\label{msigprime2}
\end{equation}
with $\sigma_{10} = \sigma/10\, {\rm km\,s^{-1}}$, and from Fig. 3 we find that Eddington factors $\mout \ga 100$ are needed to reduce $M'_{\sigma}$ to typical stellar--mass values, $\la 20\msun$, and so $\dot M \ga 2\times 10^{-7}\msun\,{\rm yr^{-1}}$. This is actually less than the likely mass transfer rate for SS433. The high Eddington factor suggests a bright ULX, but to quantify this we need a model for the collimation factor $b$. For the model suggested by King (2009) we have $b = 73\macc^{-2}$, and equation (11) of that paper gives an apparent (assumed isotropic) luminosity
\begin{equation}
L_{\rm sph} \simeq 2.4\times 10^{42}\, {\rm erg\,s^{-1}}. 
\label{ulxfeed}
\end{equation}
Using equation (\ref{unbindprime}) for $\sigma \sim 10\,{\rm km\,s^{-1}}$ and $\Delta M' \sim M$ (since the binary exchanges a large fraction of its total mass) we find 
that a system like this could in principle expel all the gas from a host galaxy with stellar mass almost $10^{10}$ times greater than the black hole.

It is not straightforward to test this observationally, since the nebula is created over a timescale $R/v_{\rm out} \sim 7R_{\rm kpc}\sigma_{10}^{-2/3}$\, Myr (where $R_{\rm kpc}$ is the nebula radius in kpc), and the ULX luminosity is likely to have varied over this timescale -- and in extreme cases disappeared altogether. In addition the ULX may be on an orbit through the galaxy taking it far away from much of the interstellar gas for most of its lifetime. In such cases the ULX would not disturb the gas significantly, and we would expect the nascent central SMBH to grow towards the usual $M_{\sigma}$ value (\ref{msig}), as appears to have happened in the dwarf galaxy RGG 118 ($\sigma \sim 30\,{\rm km\, s^{-1}}$, Baldassare et al., 2015). But is is clear that our theoretical picture shows that bright ULXs can have noticeable effects on their hosts, in line with observation.

\section{Conclusion}

We have shown that black holes supplied with mass at hyper--Eddington rates produce outflows characterized by velocities $v'$ of order $0.1 - 0.2c$ for Eddington factors $\dot m \sim 10 - 100$. The observed presence of such winds in ULXs supports the view that they are stellar--mass compact binaries in strongly super--Eddington accretion states. The photospheric temperatures of the winds are $\sim 100$\,eV, as seen in the ultraluminous supersoft sources (ULSs), again interpreted as ULXs seen `from the side'. At higher Eddington factors the winds are predicted to be slower, strengthening the connection with the highly--energetic system SS433, also seen as a ULX viewed from outside the main X--ray emission cone, but this time with photospheric emission too soft to be detected as a ULS. 

Hyper--Eddington winds have mechanical luminosities a few times the Eddington value $\led$, and can exert strong feedback on their surroundings.
For supermassive black hole (SMBH) accretion this is significantly more powerful than the usual near--Eddington (`UFO') case, and would imply $M - \sigma$ masses noticeably smaller than observed. We suggest that since the discs in SMBH accretion events are likely to be significantly warped in most cases, the central wind may often be able to blow away most of the disc mass, curtailing the hyper--Eddington phase, and so the tendency towards smaller $M - \sigma$ masses.

For stellar--mass accretion, the relatively stronger effects of hyper--Eddington feedback can leave significant imprints on host galaxies. These may be visible even when no ULX is seen (because we do not lie in the collimated beam) or when no ULS is detectable (its photospheric emission may be too soft), or simply because the hyper--accretion phase causing the damage has ended. Feedback from super--Eddington accreting binaries is likely to have important consequences for the formation and survival of small galaxies, and bear on the missing satellite and too big to fail problems (Boylan--Kolchin et al., 2011).  We shall investigate this in future papers.

\section*{Acknowledgments}
%\label{acknowledgments}
SIM acknowledges the support of the STFC consolidated grant ST/K001000/1 to the astrophysics group at the University of Leicester. We thank Ken Ohsuga for very helpful discussions, and the referee for a perceptive report.

%\section*{Bibliography}

\end{document}